\documentclass{tdp}
\usepackage{amsmath,natbib}
\usepackage{amsfonts}
\usepackage{graphicx, graphics,bm, url}
\usepackage{multicol}
\usepackage{float, xcolor}
\usepackage{fancyvrb}
\usepackage{comment}

\begin{document}

\title{Identification Risks Evaluation of Partially Synthetic Data with the \\ \texttt{IdentificationRiskCalculation} \\ R Package}
\author{Ryan Hornby and Jingchen Hu}
\address{Vassar College, Poughkeepsie, NY 12604, USA}


\TDPRunningAuthors{Ryan Hornby, Jingchen Hu}
\TDPRunningTitle{Running title}

\maketitle

\begin{abstract}
We extend a general approach to evaluating identification risk of synthesized variables in partially synthetic data. For multiple continuous synthesized variables, we introduce the use of a radius $r$ in the construction of identification risk probability of each target record, and illustrate with working examples.  We create the \texttt{IdentificationRiskCalculation} R package to aid researchers and data disseminators in performing these identification risks evaluation calculations. We demonstrate our methods through the R package with applications to a data sample from the Consumer Expenditure Surveys, and discuss the impacts on risk and data utility of 1) the choice of radius $r$, 2) the choice of synthesized variables, and 3) the choice of number of synthetic datasets. We give recommendations for statistical agencies for synthesizing and evaluating identification risk of continuous variables.
\keywords{Identification disclosure risks, Privacy protection, Synthetic data, Utility-risk trade-off}
\end{abstract}

\section{Introduction}

Data synthesis has been actively researched and practiced for microdata dissemination, where record-level data of individuals or business establishments, is collected and released to the public. Specifically, statistical agencies build statistical models on the original, confidential data, and generate records based on the model estimations \citep{Rubin1993synthetic, Little1993synthetic, RaghuReiterRubin2003JOS, ReiterRaghu2007}. These generated data, called synthetic data, undergo evaluations of utility and disclosure risks, and is released to the public when the utility level and the risk level are satisfactory \citep{Drechsler2011book}. 


Many work has been done on various utility measures and evaluations, including \citet{Woo2009JPC} and \citet{KarrKohnenOganianReiterSanil2006}, and there seems to be some consensus: statistical agencies should check global utility of synthetic data, where similarity between the original data distribution and the synthetic data distribution is evaluated (e.g. propensity scores metric \citep{Woo2009JPC}); moreover, statistical agencies should check analysis-specific utility of synthetic data, such as estimation of quantities of interest (e.g. mean, quantiles, and regression coefficients), and compare them between the original and the synthetic data through metric such as the interval overlap \citep{KarrKohnenOganianReiterSanil2006}.

By contrast, proposed approaches to identification disclosure risk evaluation in the literature can be specific to the applications, especially for continuous synthesized variables: \citet{Reiter2005CART} uses matching-based approaches based on the most frequent and average value. \citet{DomingoMateoTorra2001} and \citet{KimKarrReiter2016JOS} calculate and report the percentages of records for which the correct link is the closest, second, or third closest match. \citet{WangReiter2012} use a Bayesian approach where they calculate the average Euclidean distance that the intruder's prior would be away from the actual location data, leading to counting and reporting the number of other data points that are within a circle with the calculated average Euclidean distance. See \citet{Hu2019TDP} for an extensive review.

\citet{ReiterMitra2009} proposed a general matching-based approach to evaluating the identification risks of partially synthetic data, mainly for categorical variables. \citet{HuSavitskyWilliams2020rebds} proposed the use of radius $r$ for identification risks evaluation of one continuous synthesized variable. In this work, we extend this radius-based approach to cases where multiple continuous variables are synthesized. For each synthesized continuous variable $x_i$ of record $i$, we cast a range $R(x_i, r) = [x_i - r, x_i + r]$ around $x_i$ with radius $r$, and declare a match if a record $j$'s synthetic value $x^*_j$ falls into this range, i.e. $\mathbb{I}(x^*_j \in R(x_i, r)) = 1$. Using intuitive working examples, we illustrate how the range works for cases with multiple continuous synthesized variables. 

We create the \texttt{IdentificationRiskCalculation} R package to facilitate these identification risks evaluation calculations \citep{IR_github}. We demonstrate our proposed identification risk evaluation methods using our R package with applications to a data sample of the Consumer Expenditure Surveys (CE), published by the U.S. Bureau of Labor Statistics\footnote{For information about the CE public-use microdata (PUMD), visit: \url{https://www.bls.gov/cex/pumd.htm}}. Our data sample comes from the 3rd quartile in 2018. It includes sensitive continuous variables, such as income and expenditure, which will be synthesized and evaluated for identification disclosure risks; variable details are in Table \ref{tab:data_table}.

\begin{table}[]
    \centering
    \begin{tabular}{p{0.8in} p{0.8in} p{3in}}
    \hline Variable & Type &Description \\\hline
    Age &  Continuous & Age of reference person. \\
    Urban & Categorical & Whether this CU located in an urban or rural area. \\
    Tenure & Categorical & Housing tenure. \\
    Educ & Categorical & Education level of reference person. \\
    Expenditure & Continuous & Total expenditure last quarter. \\
    Marital & Categorical & Marital status of reference person.\\
    Income & Continuous & Total amount of family income before taxes in the last 12 months.\\\hline 
\end{tabular}
\vspace{1mm}
    \caption{Variables used from the CE data sample.}
    \label{tab:data_table}
\end{table}

In our applications, we discuss the effect of the choice of radius $r$ on identification risk, the effect of variable choices on the utility-risk trade-off, and the effect of the number of synthetic datasets on the utility-risk trade-off. In all applications, we use CART models to perform data synthesis with the default settings in the \texttt{synthpop} package \citep{synthpopJSS}.

The remainder of the article is organized as follows. In Section \ref{IR} we describe our proposed identification risk evaluation methods of synthesized variables of different data types, with definition, illustrations, and discussions of implications. Section \ref{package} presents information about the \texttt{IdentificationRiskCalculation} R package. Section \ref{apps} presents our applications to the CE data sample, with three subsections focusing the effects of radius $r$, the effects of synthesized variables, and the effects of the number of synthetic datasets. We end with a few concluding remarks in Section \ref{concluding}.

\section{Identification Risk Evaluation of Continuous Synthesized Variables}
\label{IR}

The main evaluation approach to identification risk evaluation of partially synthetic data is proposed by \citet{ReiterMitra2009}, where three summaries of identification risk measures are commonly used: 1) the expected match risk, 2) the true match rate, and 3) the false match rate. Most of applications of this approach to identification risk evaluation only consider categorical variables \citep{HuHoshino2018PSD, DrechslerHu2018, HuSavitsky2018}, with the exception of \citet{HuSavitskyWilliams2020rebds}, who considers a univariate continuous variable. 

We follow the general approach in \citet{ReiterMitra2009} and focus on the expected match risk as the identification risk. Specifically, we define the record-level identification risk, $IR_i$, as:
\begin{align}
    IR_i &= \frac{T_i}{c_i},
    \label{eq:IR_i}
\end{align}
where $c_i$ is the number of records with the highest match probability for the target record $i$. $T_i = 1$ if the target is among the $c_i$ number of matched records, and 0 otherwise. 

The calculation of $c_i$ and $T_i$ depends on the variable type of synthesized variables, which we will introduce in detail in each subsection with an illustrative example. Overall, the expression of $IR_i$ in Equation (\ref{eq:IR_i}) captures the probability of record $i$ being correctly identified, and this setup ensures that $IR_i \in [0, 1]$. We can also obtain a file-level summary of identification disclosure risk of $n$ records as below, a summary we use in our applications in Section \ref{apps}:

\begin{equation}
    IR = \sum_{i=1}^{n}IR_i.
    \label{eq:IR}
\end{equation}

\subsection{Categorical variables}
\label{IR:cat}

For categorical variables, to obtain the set of records with the highest match probability, we assume the intruder has access to a combination of known, unsynthesized variables. For simplicity, consider all known, unsynthesized variables as categorical, though our proposed method generalizes to continuous known, unsynthesized variables in a straightforward manner.

We assume the intruder knows the \emph{true confidential} values of the synthesized categorical variables of the target record $i$--it is reasonable to assume an intruder trying to identify a record in the synthetic data with the knowledge of the true confidential values of the synthesized variables. Equipped with information of the known, unsynthesized variables \emph{and} the true values of the synthesized categorical variables of record $i$, the intruder will then search for records in the synthetic data, who share the same available information with the target record $i$. Formally, we compute $c_i$ by: 
\begin{gather}
    c_i=\sum_{j=1}^n K_i(j) S_i(j),
\end{gather}
where $K_i(j)$ and $S_i(j)$ are binary indicators. $K_i(j) = 1$ if the known, unsynthesized variables of record $j$ are the same as the target record $i$, and 0 otherwise. Similarly, $S_i(j) = 1$ if the synthesized categorical variables of record $j$ are the same as the target record $i$, and 0 otherwise.

After finding $c_i$ matched records, the intruder proceeds to figure out which of the matched records is the target record $i$, and we need to evaluate how likely a correct match happens. On the one hand, if synthesized categorical variables of target record $i$ result in $S_i(i) = 1$ (i.e. $T_i = 1$), then the target record is among the matched records. In this case, the intruder will simply randomly guess which of the $c_i$ records is $i$, therefore the identification risk probability is $IR_i = 1 / c_i$. On the other hand, if synthesized categorical variables of target record $i$ result in $S_i(i) = 0$ (i.e. $T_i = 0$), then the target record is not among the matched records. In this case, the intruder will have a 0 probability of finding the true identify $i$, therefore $IR_i = 0 / c_i = 0$. 

Given this basic setup, we now turn to our identification disclosure risk evaluation of synthesized continuous variables, where we modify the approach to adapt to their continuous nature. Specifically, instead of looking for an exact match as in the categorical case, we assume matching within a distance, defined through a radius $r$. 

\subsection{One continuous variable}
\label{IR:1con}

We start with one synthesized continuous variable. As before, we assume the intruder knows the \emph{true confidential} value of the synthesized variable, $x_i$ of record $i$, when trying to find the identity of this record. We define a range $R(x_i, r) = [x_i - r, x_i + r]$ around $x_i$ with a radius $r$. We then define $c_i$, the number of records with the highest match probability for target record $i$ as:
\begin{align}
    c_i &= \sum_{j=1}^n K_i(j) \mathbb{I}(x^*_j \in R(x_i, r)),
\end{align}
where $K_i(j)$ is the binary indicator of whether record $j$ shares the same known variables as record $i$. $x^*_j$ indicates the synthetic value of record $j$. Moreover, $\mathbb{I}(\cdot)$ is a binary indicator, where $\mathbb{I}(x^*_j \in R(x_i, r)) = 1$ if $x^*_j \in [x_i - r, x_i + r]$, and 0 otherwise.

As can be seen in our setup, we do not perform an exact match for continuous $x^*_j$ as $x^*_j = x_i$. Rather, we perform matching based on a range of value of $x_i$, specifically, $[x_i - r, x_i + r]$. We recommend a percentage radius $r$ to reflect the magnitude of $x_i$. For example, $r = 20\%$ results in a range of $[x_i - 0.2x_i, x_i + 0.2x_i] = [0.8x_i, 1.2x_i]$.

We now illustrate our approach with two examples of the CE data sample, following the legends below.
\begin{table}[]
    \centering
    \begin{tabular}{l c}
        \text{True value of record }$i$ & \includegraphics[width=0.5cm]{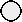} \\
        \text{Synthesized value of record }$i$ & \includegraphics[width=0.5cm]{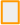}\\
        \text{Synthesized value of record }$j\neq i$ & \includegraphics[width=0.5cm]{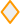} 
    \end{tabular}
\end{table}

\begin{figure}[H]
    \centering
    \includegraphics[width=\linewidth]{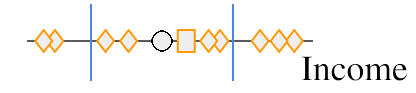}
    \caption{Case 1 - $IR_i = T_i / c_i = 1 / 5$.}
    \label{fig:1con_illustrate1}
\end{figure}

\begin{figure}[H]
    \centering
    \includegraphics[width=\linewidth]{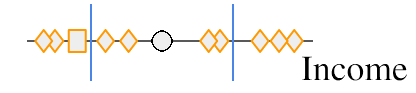}
    \caption{Case 2 - $IR_i = T_i / c_i = 0 / 4$.}
    \label{fig:1con_illustrate2}
\end{figure}

We assume the continuous variable income is synthesized. Our matching for record $i$ is based on the range $R(\text{Income}_i, r)$, which is indicated by the two vertical blue bars in the two cases in Figure \ref{fig:1con_illustrate1} and Figure \ref{fig:1con_illustrate2}. As can be seen, for one continuous variable, the range results in an interval.

\begin{multicols}{2}

\end{multicols}

For case 1 in Figure \ref{fig:1con_illustrate1}, with 5 synthetic records inside the interval, $c_i = 5$. Moreover, the synthetic income of record $i$ (the square) is in the interval, $T_i=1$, resulting in $IR_i = T_i / c_i = 1 / 5$. For case 2 in Figure \ref{fig:1con_illustrate2}, with 4 synthetic records inside the interval, $c_i = 4$. However, in this case the synthetic income of record $i$ (the square) is \emph{not} in the interval, $T_i = 0$, resulting in $IR_i = T_i / c_i = 0 / 4 = 0$.


\subsection{Two continuous variables}
\label{IR:2con}

We now move to two synthesized continuous variables, and the \emph{true confidential} values are $x_i$ and $y_i$, respectively, for record $i$. We therefore have two ranges, $R(x_i, r) = [x_i - r, x_i + r]$ and $R(y_i, r) = [y_i - r, y_i + r]$. We then define $c_i$, the number of records with the highest match probability for target record $i$ as:
\begin{gather}
    c_i = \sum_{j=1}^n K_i(j) \mathbb{I}(x^*_j \in R(x_i, r) \cap y^*_j \in R(y_i, r)),
\end{gather}
where $K_i(j)$ is the binary indicator of whether record $j$ shares the same known variables as record $i$. $x^*_j$ and $y^*_j$ denote the synthetic values of record $j$ of these two continuous variables. Moreover, $\mathbb{I}(\cdot)$ is a binary indicator, where $\mathbb{I}(x^*_j \in R(x_i, r) \cap y^*_j \in R(y_i, r) = 1$ if $x^*_j \in [x_i - r, x_i + r]$ and $y^*_j \in [y_i - r, y_i + r]$, and 0 otherwise. That is, we declare a match only when the synthetic values of \emph{both} variables of record $j$ fall into their associated ranges. We recommend a percentage radius $r$ as before, and one can use different percentage values for different continuous variables.

\begin{figure}[H]
    \centering
    \includegraphics[width=\linewidth]{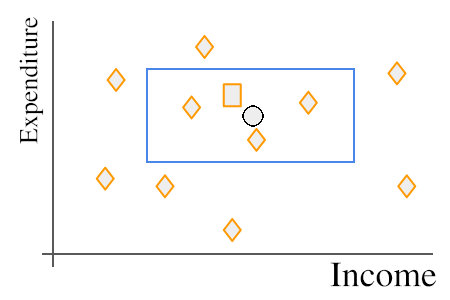}
    \caption{Case 3 - $IR_i = T_i / c_i = 1 / 4$.}
    \label{fig:2con_illustrate1}
\end{figure}

\begin{figure}[H]
    \centering
    \includegraphics[width=\linewidth]{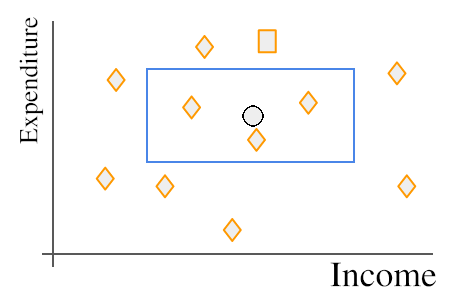}
    \caption{Case 4 - $IR_i = T_i / c_i = 0 / 3$.}
    \label{fig:2con_illustrate2}
\end{figure}

We now illustrate our approach with two examples of the CE data sample, following the same legends in Section \ref{IR:1con}. We assume income and expenditure are the two synthesized continuous variables. Therefore, our matching for record $i$ is based on the two ranges $R(\text{Income}_i, r)$ and $R(\text{Expenditure}_i, r)$, which is indicated by the blue rectangle in Figure \ref{fig:2con_illustrate1} and Figure \ref{fig:2con_illustrate2}. As can be seen, for two continuous variables, the combination of two ranges forms a rectangle.

For case 3 in Figure \ref{fig:2con_illustrate1}, with 4 synthetic records inside the rectangle, $c_i = 4$. Moreover, the synthetic income \emph{and} expenditure of record $i$ are in their corresponding ranges, so that record $i$ (the square) is \emph{in} the rectangle, $T_i = 1$, resulting in $IR_i = T_i / c_i = 1 / 4$. For case 4 in Figure \ref{fig:2con_illustrate2}, with 3 synthetic records inside the rectangle $c_i = 3$. However, in this case the synthetic income is in its range but the synthetic expenditure is not, resulting record $i$ (the square) being \emph{outside} of the rectangle. Therefore, $T_i = 0$ resulting in $IR_i = T_i / c_i = 0 / 3 = 0$.

\begin{multicols}{2}

\end{multicols}

\begin{figure}[H]
    \centering
    \includegraphics[width=0.5\linewidth]{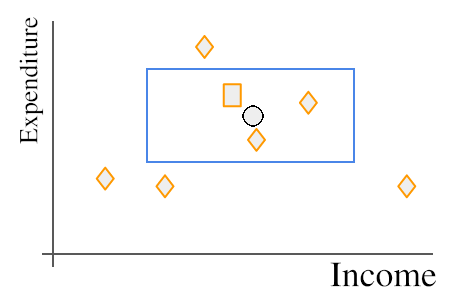}
    \caption{Case 5 - $IR_i = T_i / c_i = 1 / 3$.}
    \label{fig:2con_more_illustrate2}
\end{figure}

Our proposed identification risk evaluation methods can be extended to more than two continuous variables in a straightforward manner. For example, when working with three continuous synthesized variables, the combination of three ranges forms a cuboid. Moreover, it is worth illustrating how our identification risk evaluation setup adapts to cases where additional categorical variables are synthesized. We introduce case 5 in Figure \ref{fig:2con_more_illustrate2}, which compared to case 3 in Figure \ref{fig:2con_illustrate1}, an additional categorical variable, for example tenure from the CE sample, is synthesized.

As can be seen in the differences between case 3 and case 5, the effects of synthesizing an additional categorical variable are that fewer synthetic records remain matched overall, because we have one more variable to match with. The effects might impact both $T_i$ and $c_i$. In the illustrative example of case 5, we have $IR_i = T_i / c_i = 1 / 3$.

\section{The \texttt{IdentificationRiskCalculation} R package}
\label{package}

We have created an R package, \texttt{IdentificationRiskCalculation}, to facilitate the identification risk evaluation computation of our proposed methods \citep{IR_github}. In this section we describe how to use the function \texttt{IdentificationRisk()} and what outputs it produces. The package can be installed with the following R code:

\begin{Verbatim}[frame=single]
library(devtools)
install_github("RyanHornby/IdentificationRiskCalculation")
\end{Verbatim}

The \texttt{IdentificationRisk()} function in this R package computes the identification risks for \emph{all} records in the confidential dataset. Its outputs include matrices listing the values for $c_i$, $T_i$, and $IR_i$ for each record (row) and each synthetic dataset (column). Additionally vectors of the true and false match rates are produced, if users are interested in reporting them \citep{ReiterMitra2009}. 

Bellow is an example call of the \texttt{IdentificationRisk()} function:
\begin{Verbatim}[frame=single]
IdentificationRisk(origdata, 
                   syndata, 
                   known, 
                   syn, 
                   r,
                   percentage = true,
                   euclideanDist = false)
\end{Verbatim}

The first required argument is \texttt{origdata} which is a dataframe of the confidential dataset. The next argument is \texttt{syndata} which is a list of synthetic datasets. The next two arguments, \texttt{known} and \texttt{syn}, are both vectors. These vectors contain the names of the columns corresponding to the known and synthesized variables respectively. The final required argument is \texttt{r} which is radius to evaluate the continuous variables at. This radius can be either one value or a vector of length equal to the total number of continuous variables in the known and synthetic vectors. 

There are a couple of optional arguments that work with this radius. First is \texttt{percentage}, which by default is set to \texttt{true} such that the radius value(s) are calculated based on the confidential values. Next is \texttt{euclideanDist}, which by default is set to \texttt{false} such that each radius is treated independently leading to a rectangle in the 2D case instead of an ellipse. 


We now proceed to demonstrate the usage of this R package with CE applications in Section \ref{apps}. 

\section{Applications to the CE Data Sample}
\label{apps}
In this section, we present three applications with four scenarios to the CE data sample to demonstrate our identification risk evaluation methods, available in our R package. We use the \texttt{synthpop} R package to perform data synthesis \citep{synthpopJSS}. Specifically, we use CART synthesis models to synthesize several variables in the CE data sample, some are continuous and some are categorical. All three applications are partially synthetic data, where a subset of variables are synthesized \citep{Little1993synthetic}. This CE sample is available in our \texttt{IdentificationRiskCalculation} R package.


In all three applications, we assume the intruder knows the values of \{Age, Urban, Martial\} of each record in the CE data sample. These variables are not synthesized in any application. Note that Age is a continuous variable, and we use the percentage radius method in Section \ref{IR:1con} for its matching. Given these assumptions, we start with the following R code:

\begin{Verbatim}[frame=single]
CEdata <- IdentificationRiskCalculation::CEdata
CEdata$Urban <- as.factor(CEdata$Urban)
CEdata$Marital <- as.factor(CEdata$Marital)
CEdata$Tenure <- as.factor(CEdata$Tenure)

knownvars <- c("Age", "Urban", "Marital")
r_age <- 0.1    
\end{Verbatim}

Each of the four scenarios is associated with a set of synthesized variables. Table \ref{tab:scenarios} lists the detail of each scenario, including the nature (continuous vs categorical; ``con" stands for continuous and ``cat" stands for categorical) of each variable and the synthesis order (from left to right).

\begin{table}[h]
    \centering
    \begin{tabular}{c p{3in}}
    \hline Scenario & Synthesized variables \\\hline
    (1) & Income (con) \\
    (2) & Tenure (cat), Income (con) \\
    (3) & Expenditure (con), Income (con) \\
    (4) & Tenure (cat), Expenditure (con), Income (con) \\ \hline 
\end{tabular}
\vspace{1mm}
    \caption{Detail of the four scenarios of synthesized variables.}
    \label{tab:scenarios}
\end{table}

We provide sample R code for synthesizing with the \texttt{synthpop} package and calculating the risks with our package for each of these scenarios using the following code, starting with scenario 1:


\begin{Verbatim}[frame=single]
synvars1 <- c("Income")
syndata1 <- synthpop::syn(CEdata, 
                          m = 20, 
                          visit.sequence = synvars1)
r_income <- 0.1
riskList1 <- IdentificationRisk(origdata = CEdata, 
                                syndata = syndata1$syn, 
                                known = knownvars, 
                                syn = synvars1, 
                                r = c(r_age, r_income))
IR1 <- riskList1$exp.risk_vector
\end{Verbatim}

Next for scenario 2:
\begin{Verbatim}[frame=single]
synvars2 <- c("Tenure","Income")
syndata2 <- synthpop::syn(CEdata, 
                          m = 20, 
                          visit.sequence = synvars2)
r_income <- 0.1
riskList2 <- IdentificationRisk(origdata = CEdata, 
                                syndata = syndata2$syn, 
                                known = knownvars, 
                                syn = synvars2, 
                                r = c(r_age, r_income))
IR2 <- riskList2$exp.risk_vector
\end{Verbatim}

Now for scenario 3:
\begin{Verbatim}[frame=single]
synvars3 <- c("Expenditure","Income")
syndata3 <- synthpop::syn(CEdata,
                          m = 20, 
                          visit.sequence = synvars3)
r_income <- 0.1
r_expenditure <- 0.1
riskList3 <- IdentificationRisk(origdata = CEdata, 
                                syndata = syndata3$syn, 
                                known = knownvars,
                                syn = synvars3, 
                                r = c(r_age, r_expenditure, 
                                      r_income))
IR3 <- riskList3$exp.risk_vector
\end{Verbatim}

Finally scenario 4:
\begin{Verbatim}[frame=single]
synvars4 <- c("Tenure", "Expenditure", "Income")
syndata4 <- synthpop::syn(CEdata, 
                          m = 20, 
                          visit.sequence = synvars4)
r_income <- 0.1
r_expenditure <- 0.1
riskList4 <- IdentificationRisk(origdata = CEdata, 
                                syndata = syndata4$syn, 
                                known = knownvars, 
                                syn = synvars4,
                                r = c(r_age, r_expenditure, 
                                      r_income))
IR4 <- riskList4$exp.risk_vector
\end{Verbatim}

Note that in the all the code included in this section we use a radius of 10\%, however care should be taken in choosing a radius as we explore in Section \ref{apps:radius}.

As can be seen in the above code, to adequately explore our proposed identification risk evaluation methods and avoid the impact of high variability when simulating only 1 synthetic dataset, we simulate 20 synthetic datasets in all three applications except for the last one. In the third application in Section \ref{apps:m}, we evaluate the choice of number of synthetic datasets and its impact on identification disclosure risk and utility.



\subsection{Effect of radius $r$ on identification risk}
\label{apps:radius}

Our identification risk evaluation approach for synthesized continuous variables in Section \ref{IR:1con} and Section \ref{IR:2con} relies on the choice of radius $r$. In our first application, we explore the choice of $r$ and its impact on the identification risk results.

\begin{figure}[h]
    \centering
    \includegraphics[width=0.8\linewidth]{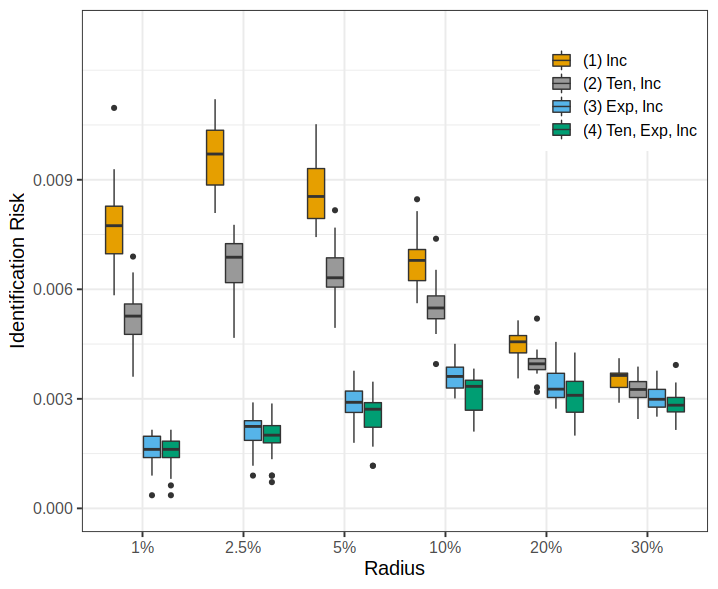}
    \caption{Effects of radius $r$ on identification risk in four scenarios.}
    \label{fig:radius}
\end{figure}

Figure \ref{fig:radius} illustrates the effect of $r = \{1\%, 2.5\%, 5\%, 10\%, 20\%, 30\%\}$ on the identification risk. We consider all four scenarios of synthesized variables listed in Table \ref{tab:scenarios}. In each scenario, we evaluate the file-level identification risk in Equation (\ref{eq:IR}) with a choice of $r$ for each of the 20 synthetic datasets. Each boxplot in Figure \ref{fig:radius} represents the distribution of the file-level identification risk across the 20 synthetic datasets under that scenario.

Figure \ref{fig:radius} shows in each scenario, there exists one $r$ value that maximizes the identification risk. For example, in scenario (1) where only income is synthesized, $r = 2.5\%$ 
is the radius value that expresses highest identification risk. After the peak as radius $r$ increases, on the one hand, records with $T_i = 1$ will remain with $T_i = 1$ (we are using an wider interval with a larger $r$, and any value in a shorter interval will remain in a wider interval when both intervals have the same center). However, as $r$ increases, $c_i$ increases since a wider interval will catch more synthetic records, making the record-level identification risk $T_i / c_i$ to decrease. On the other hand, records with $T_i = 0$ might flip to $T_i = 1$ as the interval becomes wider, although the impact of increasing $c_i$ remains. Similar discussion can be made about before the peak radius value.

Overall, in each scenario we can identify a radius $r$ that maximizes the identification risk of the simulated synthetic datasets. For scenarios (1) and (2), the maximizing $r$ is 2.5\%, while for scenarios (3) and (4), the maximizing $r$ is 10\%. We recommend the practice of finding the maximizing radius value $r$ for every scenario considered in real data applications. Moreover, we recommend that results should be reported based on the maximum identification risk, as those are most conservative: if identification risk is acceptable at its maximum, it will be acceptable at any values lower than the maximum.

For illustration purpose, we use the same value of $r$ for each continuous variable in scenarios where more than one continuous variables are synthesized (e.g. scenario (3) and (4)). In practice, statistical agencies can use different combinations of percentage radius values to explore the maximizing radius combinations, e.g. $r_{\textrm{income}}$ for income and $r_{\textrm{expenditure}}$ for expenditure in scenarios (3) and (4).


\subsection{Effects of synthesized variables on identification risk and utility}
\label{apps:moresyn}

\begin{figure}[H]
    \centering
    \includegraphics[width=0.7\linewidth]{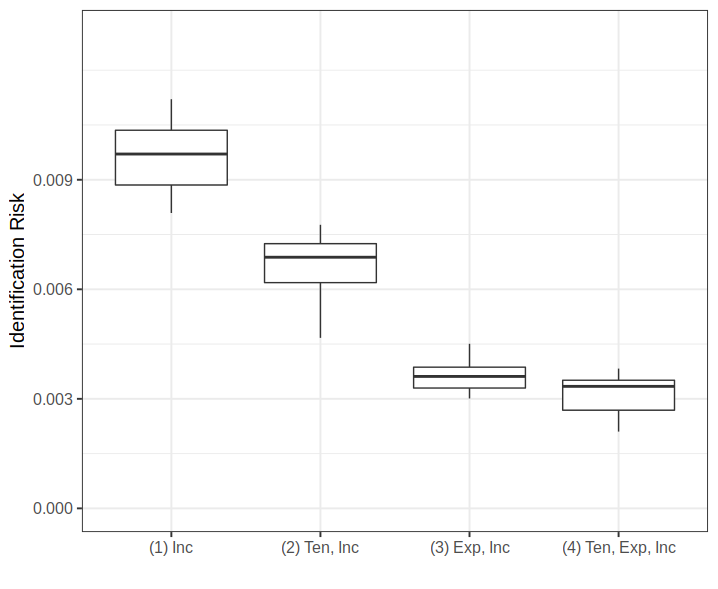}
    \caption{Identification risk of four scenarios of synthesized variables.}
    \label{fig:synth_1d}
\end{figure}

In addition to illustrating the maximizing radius $r$ for each scenario, Figure \ref{fig:radius} also shows that as more variables are synthesized, the identification risk of simulated synthetic datasets will decrease. Since different scenario reaches its maximum identification risk with a different radius $r$ value, we present Figure \ref{fig:synth_1d} where boxplots of identification risk distributions for each scenario at its maximizing radius $r$ are plotted and compared.

Figure \ref{fig:synth_1d} shows that as more variables are synthesized, identification risk decreases, a result has been demonstrated in many applications, including \citet{DrechslerHu2018}. In fact, synthesizing more variables is a typical strategy when statistical agencies need higher privacy protection.

What is interesting here is that between scenarios (2) and (3), both of which synthesize two variables, the effects of the type of variables to be synthesized on identification risk are quite different: synthesizing expenditure, a continuous variable, will provide a higher level of privacy protection compared to synthesizing tenure, a categorical variable. Note that we use the maximizing radius $r$, suggesting that synthesizing an additional continuous variable indeed provides a higher level of privacy protection. Moreover, if we compare scenarios (3) and (4), given income and expenditure are synthesized, to additionally synthesize tenure does not provide much further privacy protection. The biggest risk reduction comes with the choice of synthesizing a continuous variable.

It is well known that higher privacy protection does not come for free. As risk decreases, the usefulness of the synthetic data gets compromised, a phenomenon known as the utility-risk trade-off of synthetic data \citep{DuncanStokes2012CHANCE, DrechslerHu2018, respp_arxiv, SavitskyWilliamsHu2020ppm}. To illustrate in our applications, we consider the global utility measure of propensity scores, outlined in \citet{Woo2009JPC} and \citet{Snoke2018JRSSA}:
\begin{gather}
    U_p = \frac{1}{2n}\sum_{i=1}^{2n}\bigg(\hat{p}_i-\frac{1}{2}\bigg)^2,
\end{gather}
where $n$ is the number of records in the original dataset (and each generated synthetic dataset), and $\hat{p}_i$ is the propensity score estimated from a classification model for record $i$ belonging to the synthetic dataset. Smaller and closer-to-0 values of $U_p$ indicate high utility, since the estimated propensity score for a large number of combined records is around $1 / 2$, meaning that the chosen classification model cannot differentiate the original dataset and the synthetic dataset. In our evaluations, we use logistic regression models for classification.

\begin{figure}[H]
    \centering
    \includegraphics[width=\linewidth]{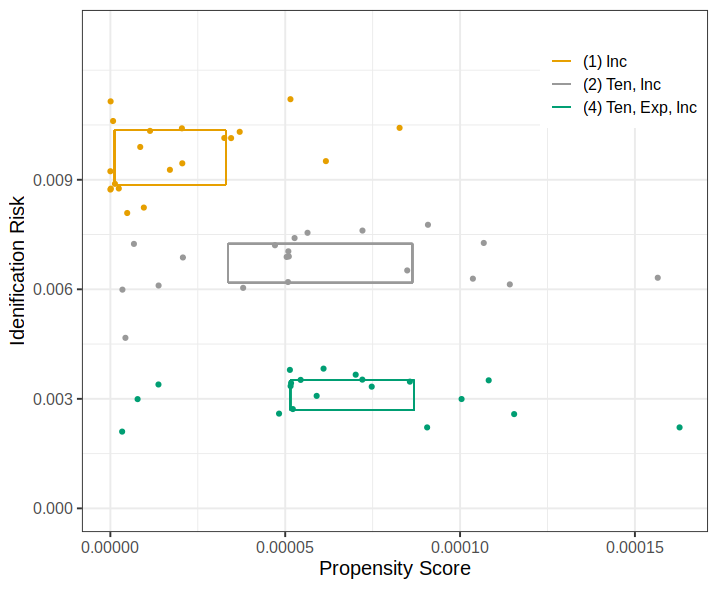}
    \caption{Utility-risk trade-off in scenarios (1), (2), and (4).}
    \label{fig:synth_2d_1}
\end{figure}

\begin{figure}[H]
    \centering
    \includegraphics[width=\linewidth]{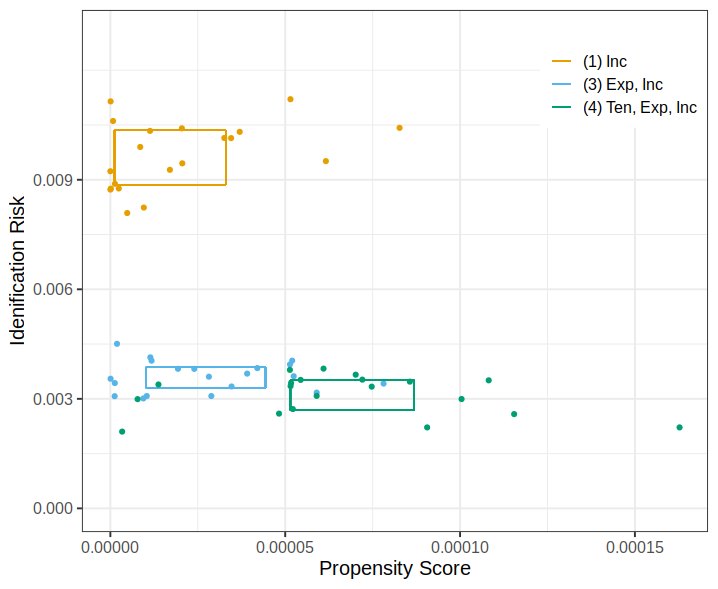}
        \caption{Utility-risk trade-off in scenarios (1), (3), and (4).}
    \label{fig:synth_2d_2}
\end{figure}

Figure \ref{fig:synth_2d_1} shows the utility-risk trade-off of scenarios (1), (2), and (4), and Figure \ref{fig:synth_2d_2} shows that of scenarios (1), (3), and (4). In each figure, we plot the identification risk and propensity score of each of the 20 synthetic datasets generated under the three scenarios. Moreover, in each scenario, we add a rectangle showing the 1st quartile and the 3rd quartile on both dimensions: utility on the x-axis and identification risk on the y-axis. The length and the height of each rectangle show the variability of each dimension across 20 synthetic datasets.

\begin{multicols}{2}

\end{multicols}

In each figure, we confirm the utility-risk trade-off: when we synthesize more variables, for example from scenario (1) to scenario (2) in Figure \ref{fig:synth_2d_1}, identification risk decreases and the utility decreases (increased propensity scores indicate lower utility). Same observation can be made for any pair of scenarios in either figure.

However, comparing scenario (2) and scenario (3) across Figure \ref{fig:synth_2d_1} and Figure \ref{fig:synth_2d_2}, we observe a case for synthesizing an additional continuous variable, as in scenario (3), over an additional categorical variable, as in scenario (2). The choice of the two continuous variables being synthesized (income and expenditure) not only expresses lower identification risk, as we have seen in Figure \ref{fig:synth_1d}, but also higher utility: the propensity score of scenario (3) are substantially lower than that of scenario (2), indicating higher utility. Moreover, the results of utility and risk express smaller variability across synthetic datasets in scenario (3), evident in the smaller size of the rectangle, on both dimensions. This suggests the robustness of our proposed radius-based matching strategy of identification risk evaluation for continuous variables.



\subsection{Effects of number of synthetic datasets on identification risk and utility}
\label{apps:m}

The synthetic data literature has traditionally recommended the release of $m > 1$ synthetic datasets with associated combining rules for valid variability estimates of quantities of interest \citep{RaghuReiterRubin2003JOS, Drechsler2011book}. More recently, recommendations of releasing $m = 1$ synthetic dataset have been made to minimize disclosure risk \citep{ReiterMitra2009, KleinSinha2015JPC, RaabNowokDibben2016JPC}. 

\begin{figure}
    \centering
    \includegraphics[width=0.7\linewidth]{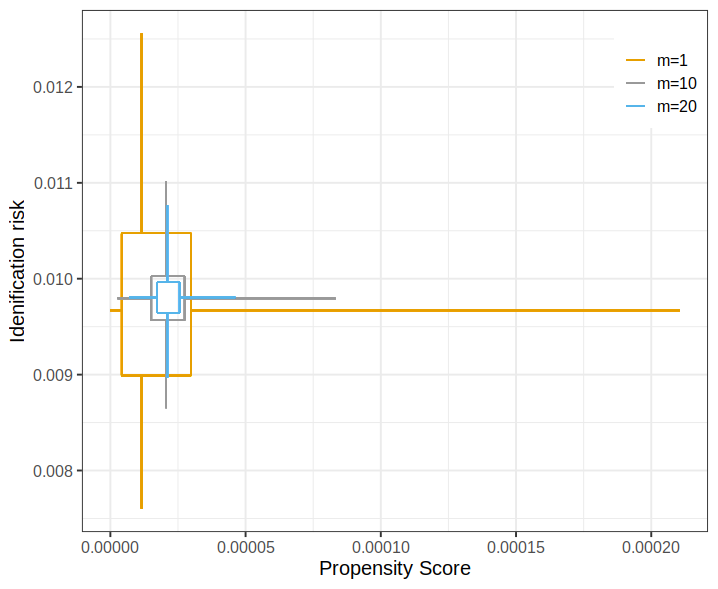}
    \caption{2D-boxplot of utility-risk trade-off in scenario (1) with different values of $m$.}
    \label{fig:m}
\end{figure}

To evaluate the effects of $m$ on identification risk and utility of continuous synthesized variables, we focus on scenario (1) where income is synthesized, and experiment with $m = \{1, 10, 20\}$. For each value of $m$, we repeat the synthetic data generation process for 1000 times, and take the average of identification risks and the average of propensity score utilities across $m$ synthetic datasets, and visualize them in Figure \ref{fig:m}. We propose the use of what we call a ``2D-boxplot" where for a given $m$, the sides of the rectangle represent the 1st quartile and the 3rd quartile across the 1000 simulations. Moreover, the lines sticking out from the rectangle start at the median values and end at the minimum and the maximum values across the 1000 simulations.

Our risk results resonate with the aforementioned works recommending $m = 1$ for higher privacy protection, as the average identification risk is the smallest for $m = 1$, and increases as $m$ increases to 10 and 20. Figure \ref{fig:m} also shows that the median identification risk is the smallest for $m = 1$, and that of $m = 10$ is slightly smaller than that of $m = 20$. 

However, what Figure \ref{fig:m} also reveals is that there is a lot more variability across the identification risk evaluations when $m = 1$, or more generally, the variability increases as $m$ decreases. As evident in the size of the three rectangles in Figure \ref{fig:m}, we could get an outcome of identification risk as small as 0.0076, and we could also get an outcome as large as 0.0126, while the minimum and the maximum values are 0.0087 and 0.0011 for $m = 10$ and 0.0090 and 0.0108 for $m = 20$. These suggest that while $m = 1$ produces the smallest average identification risk (i.e. highest privacy protection), it could severely overestimate and / or underestimate the identification risk outcome given its large spread of values. 

The utility results present a similar picture: $m = 1$ produces the smallest average propensity score, indicating highest utility level, at the cost of a larger spread. In fact, if we only rely on the average identification risk and the average propensity score, $m = 1$ is the best choice with lowest risk and highest utility. However, our use of a 2D-boxplot highlights that using $m = 1$ could present an inaccurate picture of the utility-risk trade-off of generated synthetic datasets. We also experiment with other scenarios and reach the same conclusion.


\section{Concluding Remarks}
\label{concluding}

In this article, we extend a general approach to evaluating the identification risk of multiple continuous synthesized variables. We cast a range with a radius $r$ around the true value of each synthesized continuous variable, and declare a match based on a record falling into that range. Our applications to the CE data sample recommend the practice of finding and reporting based on the maximizing radius $r$ of identification risk.

We have created and made public the \texttt{IdentificationRiskCalculation} R package for interested researchers and practitioners to implement the general approach to identification risk evaluation for categorical variables, and try out and experiment with our proposed methods for multiple continuous synthesized variables. 

We have seen that compared to synthesizing categorical variables, synthesizing continuous variables can be efficient at reducing identification risk with minimal impacts on the data utility. We also saw that generating $m > 1$ synthetic datasets achieves a more accurate picture of the utility-risk trade-of of generated synthetic datasets. We believe future research, such as using different synthesis models, different datasets, and different utility measures, are valuable.



\bibliographystyle{natbib}
\bibliography{synbib}

\end{document}